\documentstyle[12pt]{article}
\newlength{\pubnumber} 

\catcode`\@=11
\@addtoreset{equation}{section}
\def\section{\@startsection{section}{1}{\z@}{3.5ex plus 1ex minus .2ex}
 {2.3ex plus .2ex}{\large\bf}}
\def\subsection{\@startsection{subsection}{2}{\z@}{2.3ex plus .2ex}
 {2.3ex plus .2ex}{\bf}}

    \renewcommand{\baselinestretch}{1.4}
\renewcommand{\theequation}{\thesection.\arabic{equation}}

\begin{document}

\begin{titlepage}
\samepage{
\setcounter{page}{1}
\rightline{UFIFT-HEP-97-16}
\rightline{\tt hep-ph/9704363}
\rightline{March 1997}
\vfill
\begin{center}
 {\Large \bf Exotic Leptoquarks from\\
Superstring Derived Models\\}
\vfill
 {\large John K. Elwood\footnote{
   E-mail address: jelwood@phys.ufl.edu} and
	 Alon E. Faraggi\footnote{
   E-mail address: faraggi@phys.ufl.edu} \\}
\vspace{.12in}
 {\it Institute For Fundamental Theory, \\
  University of Florida, \\
  Gainesville, FL 32611 USA\\}
\end{center}
\vfill
\begin{abstract}
  {\rm
The H1 and ZEUS collaborations have recently reported
a significant excess of ${\rm e}^+p\rightarrow{e^+}{\, \rm jet}$
events at high $Q^2$.
While there exists insufficient data to conclusively determine the origin
of this excess, one possibility is that it is
due to a new leptoquark at mass scale around 200 GeV.
We examine the type of leptoquark states that exist in
superstring derived standard--like models,
and show that, while these models may contain
the standard leptoquark states which exist in Grand
Unified Theories, they also generically contain
new and exotic leptoquark states with fractional
lepton number, $\pm1/2$.
In contrast to the traditional GUT--type leptoquark
states, the couplings of the exotic leptoquarks to the
Standard Model states are generated after the breaking
of $U(1)_{B-L}$. This important feature of the
exotic leptoquark states may result in local discrete
symmetries which forbid some of the undesired leptoquark
couplings. We examine these couplings in several
models and study the phenomenological implications.
The flavor symmetries of the
superstring models are found to naturally suppress leptoquark flavor changing
processes.
}
\end{abstract}
\vfill
\smallskip}
\end{titlepage}

\setcounter{footnote}{0}

\def\beq{\begin{equation}}
\def\eeq{\end{equation}}
\def\beqn{\begin{eqnarray}}
\def\eeqn{\end{eqnarray}}
\def\AEF{A.E. Faraggi}
\def\NPB#1#2#3{{\it Nucl.\ Phys.}\/ {\bf B#1} (19#2) #3}
\def\PLB#1#2#3{{\it Phys.\ Lett.}\/ {\bf B#1} (19#2) #3}
\def\PRD#1#2#3{{\it Phys.\ Rev.}\/ {\bf D#1} (19#2) #3}
\def\PRL#1#2#3{{\it Phys.\ Rev.\ Lett.}\/ {\bf #1} (19#2) #3}
\def\PRT#1#2#3{{\it Phys.\ Rep.}\/ {\bf#1} (19#2) #3}
\def\MODA#1#2#3{{\it Mod.\ Phys.\ Lett.}\/ {\bf A#1} (19#2) #3}
\def\IJMP#1#2#3{{\it Int.\ J.\ Mod.\ Phys.}\/ {\bf A#1} (19#2) #3}
\def\nuvc#1#2#3{{\it Nuovo Cimento}\/ {\bf #1A} (#2) #3}
\def\etal{{\it et al,\/}\ }
\hyphenation{su-per-sym-met-ric non-su-per-sym-met-ric}
\hyphenation{space-time-super-sym-met-ric}
\hyphenation{mod-u-lar mod-u-lar--in-var-i-ant}


\setcounter{footnote}{0}
\section{Introduction}
The H1 \cite{hone} and ZEUS \cite{zeus}
collaborations have recently
reported an excess of events in high--$Q^2$
${\rm e}^+p\rightarrow {\rm e}^+{\, \rm jet}$ collisions.
While it is premature to conclude whether or not this
excess arises from physics beyond the Standard Model \cite{drees},
one of the possible explanations is the existence of a
leptoquark state around $O(200~{\rm GeV})$
\cite{alta,kali,bkmw,kunszt,friberg,don,blum,hew,papa}.
Leptoquark states arise generically in the context of
Grand Unified Theories, and their properties have been
discussed extensively \cite{lqreview}.
In this paper we examine the
leptoquark states that arise in superstring derived
standard--like models. These models give rise to
leptoquark states similar to those which
exist in Grand Unified Theories, as well as exotic
leptoquark states arising from the breaking
of the non--Abelian gauge symmetry to the Standard
Model gauge group at the string level,
rather than at the effective field theory level.
As a result, an important property of the exotic
``stringy'' leptoquark states is that they carry
fractional charges under the $U(1)$ generators in
the Cartan subalgebra of $SO(10)$,
$U(1)_{B-L}$ and $U(1)_{T_{3_R}}$.
Consequently, while the exotic leptoquarks carry
the usual charges under the Standard Model gauge
group, they carry ``fractional'' charge under the
$U(1)_{Z^\prime}$ symmetry\footnote{ $U(1)_{Z^\prime}$
is the combination of $U(1)_{B-L}$ and $U(1)_{T_{3_R}}$
which is orthogonal to the weak--hypercharge $U(1)_Y$.},
and therefore carry fractional
lepton number $\pm1/2$. For this reason, the
couplings of the exotic leptoquarks to the Standard Model
states are generated only after the breaking of $U(1)_{Z^\prime}$.
This is an important property of the exotic leptoquark
states, as it may give rise to local discrete symmetries \cite{lds}
that can forbid some of the undesired leptoquark couplings
not forbidden for regular leptoquarks.

In this paper we study the leptoquark states which
exist in the superstring derived standard--like models.
We first discuss how the different types of leptoquark states arise
in the superstring models. We then study the couplings
of the regular and exotic leptoquarks in several specific
models, and show that the string models under investigation
naturally give rise to symmetries which forbid some of
the undesired leptoquark couplings. For example, we find that
the flavor symmetries of the models, which arise due to
the underlying $Z_2\times Z_2$ orbifold compactification,
forbid flavor non--diagonal couplings at leading order.
We therefore arrive at the pleasing conclusion that stringy
symmetries prevent
the leptoquark states from inducing unacceptable flavor
changing interactions.

\section{The superstring standard--like models}

In this section, we give a brief overview of the superstring models
in the free fermionic formulation. It is important to note that,
although we will examine the leptoquark states in some specific
models, the types of exotic states that we describe are generic
in the free fermionic standard--like models, and similar
exotic leptoquarks may in fact
arise in general string compactifications. The purpose
of providing this brief overview is to emphasize those generic
features of the construction responsible for the standard
and exotic leptoquark states.

The superstring models that we discuss are constructed in the
free fermionic formulation \cite{fff}.
In this formulation, a model is constructed by
choosing a consistent set of boundary condition basis vectors.
The basis vectors, $b_k$, span a finite
additive group $\Xi=\sum_k{{n_k}{b_k}}$,
where $n_k=0,\cdots,{{N_{z_k}}-1}$.
The physical massless states in the Hilbert space of a given sector
$\alpha\in{\Xi}$ are obtained by acting on the vacuum with
bosonic and fermionic operators and by
applying the generalized GSO projections. The $U(1)$
charges with respect to the unbroken Cartan generators of the four
dimensional gauge group, $Q(f)$, are in one
to one correspondence with the $U(1)$
currents ${f^*}f$ for each complex fermion $f$, and are given by:
\begin{equation}
{Q(f) = {1\over 2}\alpha(f) + F(f)},
\label{u1charges}
\end{equation}
where $\alpha(f)$ is the boundary condition of the world--sheet fermion $f$
 in the sector $\alpha$, and
$F_\alpha(f)$ is a fermion number operator.

The realistic models in the free fermionic formulation are generated by
a basis of boundary condition vectors for all world--sheet fermions
\cite{revamp,fny,alr,eu,slm,gcu,lykken}.
In the models that we examine
the basis is constructed in two stages. The first stage consists
of the NAHE set \cite{slm},
which is the set of five boundary condition basis
vectors $\{{{\bf 1},S,b_1,b_2,b_3}\}$. The gauge group after the NAHE set
is $SO(10)\times SO(6)^3\times E_8$, and possesses $N=1$ space--time
supersymmetry.
The right--moving complex fermions ${\bar\psi}^{1,\cdots,5}$
produce the observable $SO(10)$ symmetry.
In addition to the gravity and gauge multiplets, the Neveu--Schwarz
sector produces six multiplets in the 10 representation
of $SO(10)$, and several $SO(10)$ singlets transforming under the
flavor $SO(6)^3$ symmetries.
The sectors $b_1$, $b_2$ and $b_3$
produce 48 spinorial 16 of $SO(10)$,
sixteen each from the sectors $b_1$, $b_2$ and $b_3$.
The free fermionic
models correspond to $Z_2\times Z_2$ orbifold models with
nontrivial background fields \cite{fm}. The NS sector
corresponds to the untwisted sector, and the sectors
$b_1$, $b_2$ and $b_3$ to the three twisted sectors
of the $Z_2\times Z_2$ orbifold model.

In the second stage of the basis construction, three
additional basis vectors are added to the NAHE set.
These three additional basis
vectors correspond to ``Wilson lines'' in the orbifold formulation.
They are needed to reduce the number of generations
to three, one each from the sectors $b_1$, $b_2$ and $b_3$.
At the same time, the additional boundary condition basis vectors
break the gauge symmetries of the NAHE set.
The $SO(10)$ symmetry is broken to one of its
subgroups, either $SU(5)\times U(1)$, $SO(6)\times SO(4)$, or
$SU(3)\times SU(2)\times U(1)_{B-L}\times U(1)_{T_{3_R}}$.
This is achieved by the following assignment of boundary conditions to the
set ${\bar\psi}^{1,\cdots,5}$:
\begin{eqnarray}
b\{{{\bar\psi}_{1\over2}^{1\cdots5}}\}&=&
\{{1\over2}~{1\over2}~{1\over2}~{1\over2}~
{1\over2}\}~\Rightarrow SU(5)\times U(1),\label{so10to64}\\
b\{{{\bar\psi}_{1\over2}^{1\cdots5}}\}&=&\{1~~1 ~~1 ~~0 ~~0\}
\Rightarrow SO(6)\times SO(4).\label{so10to51}
\end{eqnarray}
To break the $SO(10)$ symmetry to $SU(3)\times SU(2)\times
U(1)_C\times U(1)_L$ \footnote{$U(1)_C={3\over2}U(1)_{B-L};
U(1)_L=2U(1)_{T_{3_R}}.$},
both (\ref{so10to64}) and (\ref{so10to51})
are used, in two separate basis vectors.
In the superstring derived standard--like models, the three additional
basis vectors beyond the NAHE set are denoted
$\{\alpha,\beta,\gamma\}$.
The two basis vectors $\alpha$ and $\beta$ break the $SO(10)$ symmetry
to $SO(6)\times SO(4)$, while the vector $\gamma$ breaks
the $SO(10)$ symmetry to $SU(5)\times U(1)$.

In the models discussed below,
the observable gauge group after application
of the generalized GSO projections is
$SU(3)_C\times U(1)_C\times SU(2)_L\times U(1)_L
\times U(1)^3\times U(1)^n.$
The hidden $E_8$ gauge group is broken to $SU(5)\times SU(3)\times U(1)^2$,
and the flavor $SO(6)$ symmetries are broken to $U(1)^3\times U(1)^3$.
The weak hypercharge is given by
\begin{equation}
U(1)_Y={1\over 3} U(1)_C + {1\over 2} U(1)_L \ \ ,
\label{qu1y}
\end{equation}
while the orthogonal
combination is given by
\begin{equation}
U(1)_{Z^\prime}= U(1)_C - U(1)_L.
\label{quzp}
\end{equation}

The massless spectrum of the standard--like models contains
three chiral generations from the sectors which are charged under the
horizontal symmetries. Each of these consists of a 16 of
$SO(10)$, decomposed under the final $SO(10)$ subgroup as

\beqn
{e_L^c}&\equiv& ~[(1,~~{3\over2});(1,~1)]_{(~1~,~1/2~,~1)}~~;~~
{u_L^c}~\equiv ~[({\bar 3},-{1\over2});(1,-1)]_{(-2/3,1/2,-2/3)};~~~~
							\label{ulc}\\
{d_L^c}&\equiv& ~[({\bar 3},-{1\over2});(1,~1)]_{(1/3,-3/2,1/3)}~~;~~
Q~\equiv ~[(3, {1\over2});(2,~0)]_{(1/6,1/2,(2/3,-1/3))};~~~~\label{q}\\
{N_L^c}&\equiv& ~[(1,~~{3\over2});(1,-1)]_{(~0~,~5/2~,~0)}~~;~~
L~\equiv ~[(1,-{3\over2});(2,~0)]_{(-1/2,-3/2,(0,1))},~~~~\label{l}
\eeqn
where we have used the notation
\begin{equation}
[(SU(3)_C\times U(1)_C);
     (SU(2)_L\times U(1)_L)]_{(Q_Y,Q_{Z^\prime},Q_{\rm e.m.})},
\label{notation}
\end{equation}
and have written the electric charge of the two
components for the doublets.

The matter states from the NS sector and the sectors $b_1$, $b_2$
and $b_3$ transform only under the observable gauge group.
In the realistic free fermionic models, there is typically one
additional sector that produces matter states transforming
only under the observable gauge group.
Usually, this sector is a combination of
two of the vectors which extend the NAHE set.
For example, in the model of Ref. \cite{eu}, the combination
$b_1+b_2+\alpha+\beta$ produces
one pair of electroweak doublets, one pair of color triplets and five
pairs of $SO(10)$ singlets which are charged with respect to the $U(1)$
currents of the observable gauge group.
All matter states from the NS and $b_1+b_2+\alpha+\beta$ sectors
carry Standard Model charges, and are obtained by GSO projection from
the $10$ and $\overline{10}$ representation of $SO(10)$.

In addition to the states mentioned above transforming solely
under the observable gauge group, the sectors $b_j+2\gamma$
produce matter states that fall into the 16 representation of the hidden
$SO(16)$ gauge group decomposed under the final hidden gauge group.
The states from the sectors $b_j+2\gamma$ are $SO(10)$ singlets, but
are charged under the flavor $U(1)$ symmetries. The sectors
which arise from combinations of $\{b_1,b_2,b_3,\alpha,\beta,\pm\gamma\}$
produce additional massless matter states in vector--like representations.
Such states are exotic stringy states and cannot
fit into representations of the underlying $SO(10)$ symmetry group of
the NAHE set. They result from the breaking of the $SO(10)$ gauge
group at the string level via the boundary condition assignment
in Eqs. (\ref{so10to64}) and (\ref{so10to51}). These sectors give
rise to the exotic leptoquark states that we describe below.

Analysis of the fermion mass terms up to order $N=8$ reveals
the general texture of fermion mass matrices in the superstring
standard--like models \cite{fm,fh}.
The light Higgs doublets are obtained from the NS
and $b_1+b_2+\alpha+\beta$ sectors and typically consist of
${\bar h}_1$ or ${\bar h}_2$ and  $h_{45}$. The sectors
$b_1$ and $b_2$ produce the two heavy generations
and the sector $b_3$ produces the lightest generation.
This is due to the flavor $U(1)$ charges and because the
Higgs pair $h_3$ and ${\bar h}_3$ necessarily get a Planck
scale mass \cite{fm}. We adopt a notation consistent with this
numbering scheme throughout the remainder of the paper,
so that, for example, $Q_3$ represents the left--handed
up--down quark doublet.

\section{Leptoquarks from the superstring models}

There are several types of leptoquark states which arise in the
free fermionic models.
The first type are obtained from the
Neveu-Schwarz sector and from the sector $b_1+b_2+\alpha+\beta$.
At the level of the NAHE set, the NS sector gives rise to
vectorial 10 representations obtained by acting on the
vacuum with ${\bar\psi}^{1,\cdots,5}$ and ${\bar\psi}^{{1,\cdots,5}^*}$.
Thus, these states are in the $5+{\bar5}$ of $SU(5)$, and produce
the color triplets and electroweak doublets
\begin{equation}
D_{j}\equiv[(3,-1),(1,0)]_{(-1/3,-1,-1/3)}~~~~
{h}_{j}\equiv[({1},0),(2,1)]_{(1/2,-1,(1,0)) \ ,}
\label{nsstates}
\end{equation}
along with the complex conjugate representations. The $B-L$ charge
of the color triplets is $$Q_{B-L}=2/3Q_C=-2/3.$$
These states therefore carry baryon number $1/3$ and lepton number
$1$, and are standard leptoquarks, a type of leptoquark identical
to the types appearing in $SO(10)$ and $E_6$
models.  This should not be particularly surprising,
 as they are obtained from the
10 vectorial representation of $SO(10)$ by the GSO projections.
These leptoquark states from the NS sector can
appear in $SU(5)\times U(1)$, $SO(6)\times SU(4)$,
and $SU(3)\times SU(2)\times U(1)^2$ type models.
In the last two cases, there exists a superstring
doublet--triplet splitting mechanism that projects
these leptoquark states from the massless spectrum, while
the corresponding electroweak doublets remain in the light
spectrum \cite{slm}. Thus, string models can be constructed in which
all leptoquark states from the NS sector are projected
out of the massless spectrum.

The second type of leptoquark states arises from the sector
$b_1+b_2+\alpha+\beta$ \cite{halyo}. These states are similar to those
originating from the NS sector, and are obtained by acting
on the NS vacuum of the right--moving fermions, ${\bar\psi}^{1,\cdots,5}$.
The existence of leptoquark states from this sector depends as well
on the choice of boundary conditions in the basis vectors
$\{\alpha, \beta, \gamma\}$ which extend the NAHE set.
For example, in the model of Ref. \cite{eu}, one such vector--like state is
obtained
\begin{equation}
D_{45}\equiv[(3,-1),(1,0)]_{(-1/3,-1,-1/3)}~~~~
{\bar D}_{45}\equiv[({\bar3},1),(1,0)]_{(-1/3,-1,-1/3)} \ \ ,
\label{b1b2ab}
\end{equation}
while in the model of Ref. \cite{gcu}, all the leptoquark states from
this sector are projected out by the GSO projections.
The leptoquark states from this sector are identical to
those in $SO(10)$ and $E_6$ models.

There exist additional exotic leptoquark states obtained from
sectors which arise due to the breaking of $SO(10)$ to
$SU(3)\times SU(2)\times U(1)^2$.
These states come from sectors produced from combinations
of the NAHE set basis vectors and the additional basis vectors
$\{\alpha,\beta,\gamma\}$. Massless states arising from such sectors
do not fit into representations of the original $SO(10)$ symmetry,
as they carry fractional charges with
respect to the unbroken $U(1)$ generators, $U(1)_C$ and $U(1)_L$,
of the original non--Abelian $SO(10)$ Cartan subalgebra.
These fractional charges are a result of the
boundary conditions in Eqs. (\ref{so10to51}) and (\ref{so10to64}),
which break the $SO(10)$ symmetry to $SU(5)\times U(1)$ and
$SO(6)\times SO(4)$, respectively. The exotic states from these
sectors are therefore classified according to the pattern of
symmetry breaking in each sector. Sectors which contain the
vector $\alpha$ (or $\beta$, but not $\alpha+\beta$) break
the $SO(10)$ symmetry to $SO(6)\times SO(4)$, while
sectors that contain the vector $\gamma$ break the $SO(10)$
symmetry to $SU(5)\times U(1)$, and sectors containing both
$\alpha$ (or $\beta$) and $\gamma$ break the $SO(10)$ symmetry
to $SU(3)\times SU(2)\times U(1)^2$.

Sectors of the last sort, and with vacuum energy ${\rm V.E.}=-1+3/4$
in the right--moving sector, arise
frequently in the free fermionic standard--like
models. Massless states in these sectors are obtained by
acting on the vacuum with a complex fermion with $1/2$ or $-1/2$
boundary condition, and with fermionic oscillator
$1/4$. Such sectors give rise to the exotic leptoquark
states with the quantum numbers,
\beq
[(3,-{1\over4}),(1,-{1\over2})]_{(-1/3,1/4,-1/3)}~~;~~
[(\bar3,{1\over4}),(1,{1\over2})]_{(1/3,-1/4,1/3)} \ \ .
\label{uniton}
\eeq
These states have $Q_{\rm EM}=\mp1/3$, and therefore have the regular
down--type electric charge.
The $B-L$ charge and lepton number, however,
are\footnote{here $Q_L$ is the lepton number.}
\begin{equation}
Q_{B-L}=\mp{1\over6}~~~{\rm and}~~~Q_{L}=\pm{1\over2} \ .
\label{exoticqn}
\end{equation}
Such exotic states therefore carry fractional
lepton number $\pm1/2$ and, in fact, appear generically in the
free fermionic standard--like models. For example, in the model
of Ref. \cite{fny}, such states are obtained from the sectors
$b_3+\alpha\pm\gamma$ and $b_1+b_2+b_4+\alpha\pm\gamma$, in the
model of Ref. \cite{eu},  from
the sectors $b_{1,2}+b_3+\alpha\pm\gamma$ (Table \ref{matter3}),
and in the model of Ref. \cite{gcu}, from the sectors
$b_{1,2}+b_3+\beta\pm\gamma$ (Table \ref{matter4}).

We listed above the GUT--type and exotic leptoquark states
which exist in the superstring models appearing in the
literature to date. Below, we enumerate several additional types
of exotic leptoquark states that may appear in the superstring derived
models.

First we comment that in addition to the
color triplets with fractional lepton number $\pm1/2$ the
same type of sectors can also give rise to massless states
which are color singlets and which carry fractional electric
charge $\pm1/2$. This type of states may also carry fractional
lepton number $\pm1/2$. In general, this type of states
can either confined by the dynamics of a hidden
non--Abelian gauge group \cite{revamp}, or may become
superheavy by the choices of flat directions at the
string scale \cite{fc}. The analysis of ref. \cite{fc}
provides an example how the fractionally charged states
can decouple from the massless spectrum, while the
exotic leptoquarks remain in the light spectrum
at this level.

In addition to the sectors in the additive group with
$\pm1/2$ boundary conditions and with $X_R\cdot X_R=6$,
the superstring models may contain sectors with
$\pm1/2$ boundary conditions and with $X_R\cdot X_R=4$.
Whereas in the former case only one oscillator with $\nu_f=1/4$
acting on the NS vacuum was needed to get a massless
state, in the later case two such operators are needed.
The type of states arising from these sectors
depends on the boundary conditions of the complex world--sheet
fermions ${\bar\psi}^{1,\cdots,5}$. Two possibilities exist,
\beqn
b\{{{\bar\psi}_{1\over2}^{1\cdots5}}\}&=&
\{{1\over2}~~{1\over2}~~{1\over2}~~~~
{1\over2}~~~~{1\over2}\}~,\label{addtype1}\\
b\{{{\bar\psi}_{1\over2}^{1\cdots5}}\}&=&
\{{1\over2}~~{1\over2}~~{1\over2}-{1\over2}-{1\over2}\}~,\label{addtype2}
\eeqn
with the conjugate sectors, obtained by taking the opposite sign of
the $\pm1/2$ phases, producing the complex conjugate states.

Sectors with these boundary conditions and with
$X_R\cdot X_R=4$ can then give rise to
new exotic leptoquark states. In particular, as
two oscillators are now required in order to get a massless state,
these sectors may give rise to exotic diquarks, rather
than single quark states. Sectors of this type
may generally exist in the additive group. To construct an actual model
that includes a sector of this type, we simply include
a sector of the desired form in the basis vectors that define
the model. Table~\ref{model} gives an example of one such model
that gives rise to exotic diquarks from the
sector $\gamma$. We emphasize that this three
generation standard--like model should not be regarded
as a realistic model, but simply as an example that illustrates
the type of sectors giving rise to exotic diquarks, and
suggests the manner in which they may arise in concrete string models.

There are several types of exotic diquarks that may arise from
these types of string sectors. Sectors with the boundary
conditions of Eq. (\ref{addtype1}) may give rise to exotic
diquarks with quantum numbers
\begin{equation}
[(3,-{1\over4}),(2,-{1\over2})]_{(-1/3,1/4,(1/6,-5/6)}~~;~~
[(\bar3,{1\over4}),(2,{1\over2})]_{(1/3,-1/4,(-1/6,5/6)} \ \ ,
\label{exoticdq1}
\end{equation}
while sectors with the boundary
conditions of Eq. (\ref{addtype2}) may give rise to exotic
diquarks with quantum numbers
\begin{equation}
[(3,-{1\over4}),(2,{1\over2})]_{(1/6,-3/4,(2/3,-1/3)}~~;~~
[(\bar3,{1\over4}),(2,{1\over2})]_{(-1/6,3/4,(-2/3,1/3)} \ \ .
\label{exoticdq2}
\end{equation}
The states of the first type produce fractionally charged
baryons, and therefore cannot exist in a realistic low energy
spectrum, while the states of the second type are exotic diquarks
with standard $SO(10)$ weak hypercharge and ``fractional''
$U(1)_{Z^\prime}$ charge.

In general, we may anticipate the presence of additional types
 of exotic
leptoquarks and diquarks in other string models not
utilizing the NAHE set, and in which the weak hypercharge
does not have the standard $SO(10)$ embedding.
This can only be investigated in specific models.
We can, however, place some generic constraints.
For example, there is an upper limit on the weak--hypercharge
of the exotic
leptoquarks in models in which the color and weak non--Abelian
groups are obtained at level one affine lie algebras.
This follows from the constraint that the conformal dimension
of the massless states is ${\rm h}={\bar{\rm h}}=1$.
The contribution to the conformal dimension due to the
$SU(3)\times SU(2)\times U(1)^2$ charges is given by
\begin{equation}
{{C(R_3)}\over{k_3+3}}+{{C(R_2)}\over{k_2+2}}+{{Q_Y^2}\over{k_1}}\le1 \ ,
\end{equation}
where $C(R_i)$ and $k_i$ are the quadratic Casimir of the $R_i$
representation and the Kac--Moody level of the group $SU(i)$,
respectively. Models with $k_3=k_2=1$ and $k_Y=5/3$
cannot give an exotic diquark with $Q_Y\ge5/6$.
If a diquark with $Q_Y\ge5/6$ is observed
at low energies, therefore, it will exclude all level one
string models, with $SO(10)$ embedding of the weak hypercharge.
In a general string model in which the group factors are
not produced by free fermions or bosons, but rather by
higher level conformal field theories, additional types
of exotic leptoquarks and diquarks may be possible.
As we increase the level of the corresponding group factors,
states with larger charges can be obtained.
For example, at level $k=2$ with, $k_1=k_2=2$, $k_Y=10/3$ states
with $Q_Y\le11/6$ are in principle permissible.
Perturbative gauge coupling unification places strong constraints
on this possibility, however, as we discuss below.

\section{Interactions}
To study the phenomenology of the leptoquark states arising in
the superstring derived models, we examine the interaction
terms with the Standard Model states in the models of Refs. \cite{eu}
and \cite{gcu}. The cubic level and higher order non--renormalizable
terms in the superpotential are obtained by calculating correlators
between vertex operators,
$A_N\sim\langle V_1^fV_2^fV_3^b\cdots V_N^b\rangle$,
where $V_i^f$ $(V_i^b)$ are the fermionic (bosonic) vertex
operators corresponding to different fields. The non--vanishing terms
are obtained by applying the rules of Ref. \cite{kln}.
As the free fermionic standard--like models contain an
anomalous $U(1)$ symmetry, some Standard Model singlets
in the massless string spectrum must acquire a VEV near
the string scale which cancels the $D-$term equation of
the anomalous $U(1)$. In this process, some of the higher order
non--renormalizable terms become renormalizable
operators in the effective low energy field theory.

First, as the leptoquark states arise in vector--like
representations, mass terms for the leptoquark states
are expected to arise from cubic level or higher order
terms in the superpotential. For example,
in the model of Ref. \cite{eu}, we find at
cubic level the mass terms\footnote{here the notation of
Ref. \cite{eu} is used}
\begin{equation}
\xi_3D_{45}{\bar D}_{45}~ ,~ \xi_1H_{21}H_{22} \ ,
\label{masstermseu}
\end{equation}
where $\xi_1$ and $\xi_3$ are singlets under the entire four
dimensional gauge group so that their VEVs
are not constrained by the $D-$term constraints.
In this example, the leptoquark states can therefore remain light at
this level, at least in principle.
Of course, other phenomenological constraints
may require $\xi_1$ or $\xi_3$ to have non--vanishing
VEVs; one has yet to examine whether a fully realistic solution
allows either of these leptoquarks to remain light.
In the model of Ref. \cite{gcu}, which contains only
exotic leptoquarks, we find the cubic level mass
terms\footnote{here the notation of
Ref. \cite{gcu} is used}
\begin{equation}
\xi_1D_1{\bar D_1}~,~ \xi_2D_2{\bar D}_2 \ \ .
\label{masstermsgcu}
\end{equation}
Here, $\xi_1$ and $\xi_2$ are again singlets of the entire four
dimensional gauge group.
Finally, in the model of Ref. \cite{fny}, which also contains
only exotic leptoquark states, there are no cubic
level mass terms for the exotic leptoquarks. One finds in this model
a potential mass term at the quintic order\footnote{here the notation of
Ref. \cite{fny} is used}
\begin{equation}
{H_{33}H_{40}H_{31}H_{38}{\Phi_{23}}},
\label{quinticorder}
\end{equation}
where $H_{33}$ and $H_{40}$ are the exotic leptoquark states,
$H_{31}$ and $H_{38}$ are Standard Model singlets charged
under $U(1)_{Z^\prime}$, and ${\Phi_{23}}$ is an $SO(10)$ singlet.
Giving a VEV to these Standard--Model
singlets makes the exotic quark triplets heavy.

We next turn superpotential terms involving both the
leptoquarks and the Standard Model states.
The potential  leptoquark interaction terms are
\begin{eqnarray}
&&LQ{\bar D},~u_L^ce_L^cD,~d_L^cN_L^cD,\label{lqcopl}\\
&&QQD,~u_L^cd_L^c{\bar D},\label{lqcopb}\\
&&QDh \ , \label{sm1u1}\\
&&{\bar D}{\bar D}u_L^c \ , \label{sm1u2}
\end{eqnarray}
and $D{\bar D}\phi$. Severe constraints on the couplings
of the leptoquarks to the Standard Model states are imposed
by proton longevity.
If the couplings in Eq. (\ref{lqcopl}) and Eq. (\ref{lqcopb})
are not sufficiently suppressed, proton decay
is induced by leptoquark exchange.
For concreteness, we examine in detail the couplings in the models of Ref.
\cite{eu} and Ref. \cite{gcu}. In the case of the standard GUT type
leptoquarks from the Neveu--Schwarz sector and the sector
$b_1+b_2+\alpha+\beta$, such dangerous couplings are indeed
expected.
In general, if massless triplets from the Neveu--Schwarz sector
or the sector $b_1+b_2+\alpha+\beta$ exist in the massless spectrum,
then the terms in Eqs. (\ref{lqcopl},\ref{lqcopb})
are obtained either at the cubic
level of the superpotential, or from higher order nonrenormalizable
terms. For example, in the model of Table~\ref{modelwithexoticdiquarks}
(the massless
spectrum and quantum numbers are given in Ref. \cite{slm}),
we obtain at the cubic level,
\beqn
&&
{u_{1}^c}{e_{1}^c}{D}_1,~{d_{1}^c}{N_{1}^c}{D}_1,~
{u_{2}^c}{e_{2}^c}{D}_2,~{d_{1}^c}{N_{2}^c}{D}_1,\nonumber\\
&&
{{D_1}{\bar D}_2{\bar\Phi}_{12}},~
{\bar D}_1{D}_2{\Phi}_{12} \ ,
\label{dnscoupling}
\eeqn
while in the model of Ref. \cite{eu} we obtain at the quartic order,
\beqn
&&u_1^ce_1^c{D}_{45}{\bar\Phi}^-_1~,~
  u_2^ce_2^cD_{45}{\bar\Phi}^+_2~,~
  d_1^cN_1^cD_{45}\Phi_1^+~,~
  d_2^cN_2^cD_{45}{\bar\Phi}_2^-~,\nonumber\\
&&Q_1Q_1D_{45}\Phi_1^+~,~
  Q_2Q_2D_{45}{\bar\Phi}_2^-~.~
\label{d45coupling}
\eeqn
At higher orders, additional terms will appear.
We observe that leptoquark states from the NS sector, or from a sector
of the type of $b_1+b_2+\alpha+\beta$, will generally have the
undesirable couplings with the Standard Model states. Such couplings
are induced from higher order terms by the VEVs of the
Standard Model singlets which cancel the anomalous $U(1)$ $D$--term
equation. We therefore anticipate that a
leptoquark from either of these sectors
is not likely to provide a realistic possibility for a light
leptoquark state.

We turn now to the exotic leptoquark states.
Because of the fractional lepton number of
these exotic leptoquarks, direct couplings to the
Standard Model states are impossible. Coupling to
the Standard Model states can occur only
through higher order terms containing a Standard
Model singlet field with fractional lepton number $\pm1/2$.
Such additional fields generally exist in superstring
standard--like models. The coupling of the exotic leptoquarks
then depends on these extra fields in specific models,
together with a pattern of VEVs which preserves supersymmetry
at the Planck scale. The important aspect of the
fractional charges of the exotic leptoquarks is
that they may give rise to residual local discrete symmetries
that forbid the dangerous coupling of the exotic leptoquarks
to the Standard Model states.

In the context of the model of Ref. \cite{gcu}, the coupling of the exotic
leptoquarks to the Standard Model states has been examined
in detail in Refs. \cite{ccf,lds}. It was shown that, in
this model, the interaction terms of the exotic leptoquarks
with the Standard Model states vanish to all orders of
nonrenormalizable terms for the following reason.
The interaction terms take the forms $f_if_jD\phi^n$ and $f_iDD\phi^n$,
where $f_i$ and $f_j$ are the Standard Model states from the sectors
$b_1$, $b_2$ and $b_3$, and $D$ represents the exotic leptoquark.
The product of fields, $\phi^n$,
is a product of Standard Model singlets that ensures
invariance of the interaction terms under all $U(1)$ symmetries
and the string selection rules. If all the fields $\phi$ in the
string $\phi^n$ get VEVs, then the coefficients of the operators
in Eqs. (\ref{lqcopb}) and (\ref{lqcopl}) are of the order $(\phi/M)^n$, where
$M\sim10^{18}$GeV.
Because of the fractional charge of the exotic leptoquarks
under $U(1)_{Z^\prime}$, none of the interaction terms
in Eqs. (\ref{lqcopb}) and (\ref{lqcopl}) are invariant
under $U(1)_{Z^\prime}$.
The total $U(1)_{Z^\prime}$ charge of each of these interaction terms
is a multiple of $\pm(2n+1)/4$. Thus, for these terms to be allowed,
the string $\phi^n$ must break $U(1)_{Z^\prime}$ and must
have a total $U(1)_{Z^\prime}$ charge in multiple of $\pm(2n+1)/4$.
The string of Standard Model singlets must therefore contain a
field which carries fractional $U(1)_{Z^\prime}$ charge
$\pm (2n+1)/4$. In the model of Ref. \cite{gcu}, the only
Standard Model singlets with fractional $U(1)_{Z^\prime}$
charge transform as triplets of the hidden $SU(3)_H$
gauge group. As these fields transform in vector--like
representations, invariance under the symmetries of the
hidden sector guarantees that there is a residual $Z_4$
discrete symmetry which forbids the coupling of the
exotic leptoquarks to the Standard Model states to
all orders of nonrenormalizable terms. While this
symmetry ensures that the exotic leptoquark states
do not cause problems with proton decay, it also
forbids their generation at ${\rm e}^\pm p$ colliders.

Next we turn to the model of Ref. \cite{eu}.
In this model, we find at the quartic order of the
superpotential
\beqn
&&u_2^cd_2^cH_{21}H_{26}~,~
              Q_2L_2H_{21}H_{26}~,
\label{m278n4}
\eeqn
at the quintic order,
\beqn
&&L_3Q_3H_{21}H_{18}\Phi_{45}~,~L_3Q_3H_{21}H_{24}\xi_2~,~
  d^c_3u^c_3H_{21}H_{18}\Phi_{45}~,~d^c_3u^c_3H_{21}H_{24}\xi_2~,
\label{m278n5}
\eeqn
and at order $N=6$ we find for example,
\beqn
&&L_1Q_1H_{21}H_{24}{\bar\Phi}_{13}\xi_1~~,~~
  d^c_1u^c_1H_{21}H_{24}{\bar\Phi}_{13}\xi_1\nonumber\\
&&Q_1Q_1H_{22}H_{17}\phi_1^+\xi_1~~,~~
  d^c_1N^c_1H_{22}H_{17}{\Phi}_{1}^+\xi_1\nonumber\\
&&u^c_1e^c_1H_{22}H_{17}\Phi_1^-\xi_1~~,~~
  u^c_2e^c_2H_{22}H_{17}\Phi_2^+\xi_1\nonumber\\
&&Q_2Q_2H_{22}H_{23}{\bar\Phi}_2^-\Phi_{45}~~,~~
  d^c_2N^c_2H_{22}H_{23}{\bar\Phi}_{2}^-\Phi_{45} \ \ ,
\label{m278n6}
\eeqn
plus additional terms of the generic form
$f_if_iHH({{\partial W_3}/\partial\eta_i})$ which vanish
by the cubic level $F$--flatness constraints.
The important lesson to draw from this model
is that couplings of the exotic leptoquarks
to Standard Model states are generated from
nonrenormalizable terms by VEVs which break
the $U(1)_{Z^\prime}$ gauge group. Another
observation is that the exotic leptoquark
couplings are flavor diagonal.

\section{Phenomenology}
In this section, we discuss the phenomenological implications
of the standard and exotic leptoquark states appearing in the superstring
derived standard--like models.  The couplings of exotic leptoquarks
to the standard model states are typically quite constrained by low
energy phenomenology.

We make several simple observations with regard to interaction of the
exotic leptoquarks with the
Standard Model states. Although our observations
are made primarily for the model of Ref. \cite{eu},
they are in fact much more general.

The first comment concerns leptoquark induced
proton decay. Here we note that in the model of Ref. \cite{gcu}
the problem is solved entirely. This model contains
only exotic leptoquarks, since the ``regular'' leptoquark states from the
Neveu--Schwarz and $b_1+b_2+\alpha+\beta$ sectors are
removed by the GSO projections. Second, the spectrum,
charges, and symmetries are such that all interaction terms
of the exotic leptoquarks vanish identically. The exotic leptoquark
states therefore lead to no conflict with
proton lifetime constraints.
Clearly, however, the exotic leptoquark
states can neither account for the anomalous
HERA events.

We next turn to the model of Ref. \cite{eu}. Here,
all color triplets from the NS sector are removed
by GSO projection. The model does contain
one ``regular'' leptoquark from the sector $b_1+b_2+\alpha+\beta$
and one exotic leptoquark state from the sector
$b_{1,2}+b_3+\alpha\pm\gamma$, however.  We expect that the
``regular'' leptoquark states do have interaction terms with
the Standard Model states appearing at successive
orders. Interaction terms of the states from the sector
$b_1+b_2+\alpha+\beta$ with the Standard Model states
do not appear at the cubic level of the superpotential
because of the flavor symmetries, but may arise in
higher order non--renormalizable terms.
Although the higher order terms are expected to be suppressed
by powers of $(\langle\phi\rangle/M)^n$, this suppression,
in general, cannot make the dangerous couplings sufficiently small.
For this reason, it is not expected that such ``regular''
leptoquark states can be interpreted as light leptoquarks.

Next, we study the couplings of the exotic leptoquark
states in the model of Ref. \cite{eu}.
We note that such couplings generally arise from
higher order terms in this model. A novel feature of the exotic
leptoquark couplings is that such couplings can arise
only due to a VEV breaking the $U(1)_{B-L}$ symmetry.
Thus, the magnitude of the coupling of the exotic
leptoquarks to the Standard Model states is tied to
the $U(1)_{B-L}$ scale. This is a welcome
feature, as the rate of proton decay inducing
processes can be sufficiently small, even for a light
leptoquark, coupled as it is to the scale of $U(1)_{Z^\prime}$
breaking. Furthermore, upon examining the couplings in Eqs.
(\ref{m278n4},\ref{m278n5},\ref{m278n6}),
we note that the induced couplings depend
on the specific choice of fields that breaks the $U(1)_{B-L}$
symmetry. Therefore, for a specific pattern of such
VEVs compatible with the anomalous $U(1)$
D--term cancelation mechanism, it is generally possible to
allow the $U(1)_{B-L}$ breaking scale to occur at
a relatively high scale, while utilizing the freedom
in the choice of fields along with the flavor symmetries
to avoid conflict with the proton lifetime.
The question still remaining, however,
concerns the possibility of suppressing proton decay while
allowing for a large lepton or
baryon number violating coupling, but not for both.
Although this does not happen in the
model of Ref. \cite{eu} where, as can be seen from a quick examination of
Eq. (\ref{m278n5}), the product $<H_{18}\Phi_{45}>$ determines the magnitude
of both the B and L violating couplings, we claim that
such a situation may in general be possible.  As evidence to
support our claim, we note that the superstring
standard--like models occasionally give rise to
custodial symmetries that distinguish between
the lepton violating and baryon violating operators \cite{custodial}.
In summary, while the models of both Refs. \cite{gcu}
and \cite{eu} can sufficiently suppress
the couplings of the exotic leptoquarks and avoid problems
with proton decay, neither of these models allows a large
lepton number violating coupling of the exotic leptoquark
while suppressing the baryon violating coupling. We anticipate, however,
that there exists a slight modification,
or perhaps some synthesis of these models, that admits such
a possibility.

We next discuss the suppression of flavor violating couplings
in the model. We note that the couplings of the regular and
exotic leptoquark states are flavor diagonal in the first few
orders. Flavor mixing terms are therefore suppressed by several
powers of $\langle\phi\rangle/M$. This suppression
in fact arises due to the flavor symmetries of the superstring
derived models, and is a direct consequence of the fact that the
Standard Model
states from the sectors $b_j$ each carry charges with respect
to ($U(1)_{L_{j}}~;~U(1)_{L_{j+3}}$) and ($U(1)_{R_{j}}~;~U(1)_{R_{j+3}}$),
$(j=1,2,3)$. Thus, the states from each sector $b_j$ are charged
with respect to different pairs of $U(1)$ symmetries. This charge
assignment is a reflection of the underlying $Z_2\times Z_2$
orbifold compactification in which each of the twisted sectors
lies along an orthogonal plane. The states from the NS sector,
the sector $b_1+b_2+\alpha+\beta$, and the sectors which give
rise to the exotic leptoquarks, however, are neutral with respect
to $U(1)_{L_{j+3}}$ and $U(1)_{R_{j+3}}$. In order
to form a gauge invariant flavor mixing leptoquark coupling,
therefore,
we need to utilize additional fields with half integral charges
with respect to $U(1)_{L_{j+3}}$ and $U(1)_{R_{j+3}}$.
In the free fermionic models which are based on the NAHE set,
the only available fields are those from the sector $b_j+2\gamma$.
Furthermore, since these sectors preserve also the underlying
$Z_2\times Z_2$ orbifold structure, a potential mixing term
must contain at least two such fields from two different sectors.
This is the reason for the suppression of the
flavor mixing terms. Indeed, in the model of Ref. \cite{eu}
we find that two such terms for
the exotic leptoquarks first appear at order $N=7$,
\begin{equation}
d_3N_2H_{21}H_{17}\phi_{45}{\bar V}_2V_3~~~~~~~~
d_2N_3H_{21}H_{17}\phi_{45}{\bar V}_2V_3
\end{equation}
At higher orders, $N=8,...$ additional terms of this type are
expected to appear. Note that, in addition to the suppression
by the VEV which breaks $U(1)_{B-L}$, these terms have
an additional $(\langle\phi\rangle/M)^3$ suppression factor.
We conclude that the non--diagonal leptoquark couplings
are naturally suppressed in free fermionic models in which
each of the generations is obtained from a different twisted
sector of the $Z_2\times Z_2$ orbifold.

We now comment on the possibility of the existence of
exotic leptoquarks that preserve the family numbers.
First of all, due to the suppression of the
successive orders of nonrenormalizable terms,
we generally do not expect these couplings to be in conflict with
experimental constraints. As we discuss below, the
experimental constraints typically require that
the product of two operators be smaller than some
phenomenological limit. As in the string, the separate
operators arise a different orders. It is in fact
very natural that, while one of the operators is
relatively large, the other is sufficiently
suppressed so as to avoid conflict with observations.

Let us however discuss briefly the interesting possibility of
producing several exotic leptoquark states which carry
a family number.  In the model of Ref. \cite{eu}, this is not the case,
as there exists a single exotic leptoquark pair that couples to
all three generations. However, again we anticipate
that such a model may exist for the following reason.
In the superstring derived models, a combination
of the sectors $\alpha$, $\beta$ and $\gamma$ occasionally gives
rise to additional space--time vector bosons which enhance
the gauge group. This combination, when added to the vectors
$b_j$ $(j=1,2,3)$, gives rise to the sectors that produce
the additional states that must fill the representations
of the enhanced symmetry. Thus, this situation arises
because the vector combination which enhances the gauge
symmetry preserves the symmetry of the NAHE set.
In the vector combination that enhances the gauge symmetry,
the left--moving vacuum vanishes. Thus, if it is possible
to construct a model with vector combination $X$, with
$X_L\cdot X_L=4$, and whose addition to each of the
basis vectors $b_1$, $b_2$ and $b_3$ produces a vacuum
with $(b_j+X)_R\cdot (b_j+X)_R=6$ for $(j=1,2,3)$, then
this type of model would produce exotic leptoquarks from
each sector $b_j+X$ that couple diagonally
to the states from the sectors $b_j$. Again, we
expect that such a model may exist. It is
expected, however, that if such a model exists, higher order terms
will mix the family leptoquarks with the different families.
Such higher order terms are likely sufficiently suppressed to
avoid conflict with experimental constraints, however.

Let us consider a model which
contains the exotic leptoquark couplings of Ref. \cite{eu}, but lacks
couplings between the regular leptoquarks and the standard model states.
This property may arise, for instance, from additional $U(1)$ symmetries
in the effective theory, and is a generic possibility in such superstring
derived standard--like models.  In such a situation, we need concern
ourselves only with the couplings of Eq. (\ref{m278n4}) at lowest order
in the superpotential.   We expect that the standard model singlet
$H_{26}$ will develop a VEV somewhat below the string scale, and define
the dimensionless parameter $x_{26}$ appropriately:
\beqn
&&x_{26}={<H_{26}> \over M_{string}} \ \ .
\label{pheno1}
\eeqn
Even though the operators of Eq. (\ref{m278n4}) couple only to second
generation quarks and leptons, proton decay occurring at one loop level
places stringent bounds on their couplings.  Denoting their
couplings respectively as $\lambda_{22}^{''}$ and $\lambda_{22}^{'}$, one
finds for leptoquark masses below 1TeV the constraint \cite{sv}
\beqn
&&|\lambda_{22}^{''} \,  \lambda_{22}^{'} \,  x_{26}^2| < 10^{-9} \ \ .
\eeqn
In the absence of the lepton number violating coupling $\lambda_{22}^{'}$,
cosmological arguments place strong upper limits on $\lambda_{22}^{''}$
\cite{bouq},
\beqn
&&|\lambda_{22}^{''} \,  x_{26}| < 10^{-7} \ \ ,
\eeqn
although there is some suggestion of model dependence to these
determinations \cite{drein}.  Conversely, in the absence of
$\lambda_{22}^{''}$, deep inelastic experiments involving muon neutrinos demand
\cite{bgh}
\beqn
&&|\lambda_{22}^{'} \,  x_{26}| < 0.22 \, (M_{LQ}/100 GeV) \ \ ,
\eeqn
where $M_{LQ}$ is the mass of the leptoquark.  Due to the appearance of
$x_{26}$, which could be significantly less than unity, these relations are
satisfied much more naturally for the case of exotic leptoquarks than they
are for their more traditional integer lepton number cousins.

In the more general case of a model possessing generic lepton number violating
interactions
\beqn
&&\lambda_{ij}^{'} L_i Q_j H \phi^n_a(i,j) \ \ ,
\label{genlambda}
\eeqn
but no baryon number violating interactions, the phenomenological constraints
on the couplings $\lambda_{ij}^{'}$ are listed below in Table~\ref{lqtab}
\cite{bgh} \cite{dbc}.  In Eq.
(\ref{genlambda}), $H$ is a generic exotic leptoquark, and $\phi^n_a(i,j)$
is a string of $n$ standard model singlets necessary to give the appropriate
charge to the composite operator.  This product of singlets develops
expectation value $x^n_a(i,j)$ in the low energy effective theory.

\section{Gauge Coupling Unification}

In this section, we comment on the effect of
low energy leptoquarks and diquarks on gauge coupling unification.
It is well known that weakly coupled heterotic
string theory predicts unification of the
gauge couplings at a scale of the order \cite{scop}
\beq
   M_{\rm string}~ \approx ~ g_{\rm string}\,\times\,5\,\times\,10^{17}~
         {\rm GeV}~.
\label{Mstringvalue}
\eeq
where $g_{\rm string}\approx0.8$ at the unification scale.
If one assumes that the matter content above the
electroweak scale consist only of the MSSM states,
then the couplings are seen to intersect at a scale
of the order
\beq
   M_{\rm MSSM}~ \approx ~ 2\,\times\,10^{16}~
         {\rm GeV}~.
\label{MMSSMvalue}
\eeq
Thus, approximately a factor of 20 separates the two scales.
A priori one would expect that, in extrapolation of
the couplings over fifteen orders of magnitude, this small
discrepancy would have many possible resolutions.
Surprisingly, however, the problem is not easily
resolved.
In Ref. \cite{df,jmr}
a detailed analysis of string scale gauge coupling unification
was performed. All possible perturbative corrections
to the gauge couplings were taken into account,
including string threshold corrections, light SUSY thresholds,
enhanced intermediate gauge symmetry, modified weak
hypercharge normalizations and intermediate matter thresholds.
It was shown that only the existence of intermediate
matter thresholds, beyond the MSSM spectrum, can potentially
resolve the problem.
Existence of additional color triplets in the desert,
between the electroweak scale and the Planck scale,
has in fact been proposed for some time as a possible
solution to the problem of string scale gauge coupling
unification \cite{price,gcu}. Possible numerical scenarios
were presented in Ref. \cite{df}, and include the possibility
of having a leptoquark with weak hypercharge $-1/3$ near
the experimental limit, provided that additional color and weak
thresholds exist at a higher scales. Thus, the existence of
light leptoquarks with the appropriate weak hypercharge
assignment is desired from the point of view of
superstring unification.

We now utilize the one--loop renormalization group equations
to examine the effect of exotic intermediate states on unification,
and, in particular, to illuminate some of the potential problems
associated with having such states present in a theory.
The constraint of perturbative gauge coupling unification, in
tandem with the precision data on $\sin^2 \theta_W (M_Z)$ and
$\alpha_s (M_Z)$, for instance,
places significant restrictions on the masses
and charges of some diquark states.
For example, requiring that the $U(1)_Y=7/6$
diquarks present in some models \cite{bkmw}
not destroy unification restricts their masses to lie
quite close to the unification scale, not less than $0.6 M_X$ with
two representations of diquarks, and not less than $0.2 M_X$ with
one (Note, however, that the stringy exotic diquarks mentioned in
Eq. (\ref{exoticdq2}) above have the much more modest $U(1)_Y$ charges
${1 \over 6}$ and ${1 \over 3}$, respectively, and are therefore not so
restricted) .  Conversely, taking $k_2=k_3=1$, we find that diquarks
with masses in the TeV range are completely excluded by unification
for {\it any} value of $k_1$.  Neither does allowing higher levels for
$k_2$ and $k_3$ alleviate the problem.  In fact, for all allowed values of
$k_1$, $k_2$, and $k_3$, the presence of $U(1)_Y=7/6$ diquarks at the
TeV scale is completely inconsistent with gauge coupling unification.
At any rate, such comments should
only be viewed as indicative of the problem, as the enormous contribution
of these diquarks to the running of $\alpha_1$ actually causes it to
encounter its Landau pole in the desert.  With two representations of
such diquarks, this occurs near $10^{12}$ GeV, while for one representation,
the catastrophe is delayed until $10^{16}$ GeV.  Each of these cases is
of course inconsistent with unification.

\section{Conclusions}

Recent HERA data shows deviation from the Standard Model
expectations. As statistics for
the HERA data are still minimal, it is clearly premature
to conclude whether this anomaly is a signal of new physics or not.
Nevertheless, a possible explanation for the
excess of events is the presence of a new leptoquark state around 200 GeV.

In this paper, we studied the different types of
leptoquark states appearing in superstring
derived models. String models often give rise
to leptoquark and diquark states that are similar
to those that exist in GUTs. However, a notable
difference is that, in string models, leptoquark
states can exist without the need for an enhanced
non--Abelian gauge symmetry. Also interesting is the fact
that the spectrum of string models
is quite constrained. For instance, the existence
of a diquark with $Q_Y>5/6$ will exclude all level one
superstring models with $SO(10)$ embedding of the weak hypercharge.
More interestingly, however, we
have shown that superstring models generically give rise
to exotic leptoquark states that lack a
standard GUT correspondence. These exotic
leptoquarks arise due to the breaking of the
non--Abelian gauge symmetries at the string
level rather than at the level of the effective
four dimensional field theory. Moreover,
such exotic stringy leptoquarks possess interesting
properties. For example, their couplings
to the Standard Model states are generated
only after the breaking of the $U(1)_{Z^\prime}$
gauge symmetry.  Furthermore, flavor symmetries
that arise from the string models provide
sufficient suppression to avoid conflict with
experimental data. Finally, an exotic leptoquark
at 200 GeV is desirable from the perspective of gauge
coupling unification.  Its presence, along with that of
additional color and weak thresholds at higher energies,
can resolve the string--scale gauge coupling unification
problem.


It is a pleasure to thank Claudio Coriano for valuable discussions.
This work was supported in part by DOE Grant
No.\ DE-FG-0586ER40272.

\bibliographystyle{unsrt}

\vfill
\eject


\textwidth=7.5in
\oddsidemargin=-18mm
\topmargin=-5mm
\renewcommand{\baselinestretch}{1.3}
\renewcommand{\theequation}{\arabic{equation}}
\pagestyle{empty}
\begin{table}
\begin{eqnarray*}
&\begin{tabular}{|c|c|ccc|cccccccc|cccccccc|}
\hline
 ~ & $\psi^\mu$ & $\chi^{12}$ & $\chi^{34}$ & $\chi^{56}$ &
        $\overline{\psi}^{1} $ &
        $\overline{\psi}^{2} $ &
        $\overline{\psi}^{3} $ &
        $\overline{\psi}^{4} $ &
        $\overline{\psi}^{5} $ &
        $\overline{\eta}^1 $&
        $\overline{\eta}^2 $&
        $\overline{\eta}^3 $&
        $\overline{\phi}^{1} $ &
        $\overline{\phi}^{2} $ &
        $\overline{\phi}^{3} $ &
        $\overline{\phi}^{4} $ &
        $\overline{\phi}^{5} $ &
        $\overline{\phi}^{6} $ &
        $\overline{\phi}^{7} $ &
        $\overline{\phi}^{8} $ \\
\hline
   $\alpha$ & 1  & 1 & 0 & 0    & 1 & 1 & 1 & 0 & 0 & 1 & 0 & 1 &
                                  1 & 1 & 1 & 1 & 0 & 0 & 0 & 0 \\
   $\beta$  & 1  & 0 & 1 & 0    & 1 & 1 & 1 & 0 & 0 & 0 & 1 & 1 &
                                  1 & 1 & 1 & 1 & 0 & 0 & 0 & 0 \\
   $\gamma$ & 1  & 0 & 0 & 1    & ${1\over 2}$ & ${1\over 2}$ & ${1\over 2}$ &
                 ${1\over 2}$ & ${1\over 2}$ & ${1\over 2}$ & ${1\over 2}$ &
                 ${1\over 2}$ & ${1\over 2}$ & 0 & 1 & 1 & ${1\over 2}$ &
                 ${1\over 2}$ & ${1\over 2}$ & 0 \\
\hline
\end{tabular}
   \nonumber\\
&  ~ \nonumber\\
&  ~ \nonumber\\
&\begin{tabular}{|c|cccc|cccc|cccc|}
\hline
 ~&     $y^3{\overline y}^3$ & $y^4\overline y^4$ & $y^5\overline y^5$ &
        $y^6\overline y^6$ & $y^1\overline y^1$ & $y^2\overline y^2$ &
        $\omega^5\overline \omega^5$ & $\omega^6\overline\omega^6$ &
        $\omega^2\omega^3$ & $\omega^1\overline\omega^1$ &
        $\omega^4\overline\omega^4$ & $\overline\omega^2\overline\omega^3$ \\
\hline
       $\alpha$   & 1 & 0 & 0 & 1    & 0 & 0 & 1 & 0    & 0 & 0 & 1 & 1 \\
       $\beta$    & 0 & 0 & 0 & 1    & 0 & 1 & 1 & 0    & 0 & 1 & 0 & 1 \\
       $\gamma$   & 1 & 1 & 0 & 0    & 1 & 1 & 0 & 0    & 0 & 0 & 0 & 1 \\
\hline
\end{tabular}
\end{eqnarray*}
\caption{A three generation $SU(3)\times SU(2)\times U(1)^2$ model.
The choice of generalized GSO coefficients is:
$c{{b_1,b_2,\alpha,\beta,\gamma}\choose \alpha}=
-c{b_2\choose \alpha}=
 c{{{\bf 1},b_j,\gamma}\choose \beta}=
-c{\gamma\choose {{\bf1},b_1,b_2}}=
 c{\gamma\choose b_3}=-1~$
$(j=1,2,3)$, with the others
specified by modular invariance and space--time supersymmetry. This model
contains two leptoquarks pairs, ${D_1,{\bar D}_1,D_2,{\bar D}_2}$, from the
Neveu--Schwarz sector.}
\label{model}
\end{table}

\textwidth=7.5in
\oddsidemargin=-18mm
\topmargin=-5mm
\renewcommand{\baselinestretch}{1.3}
\pagestyle{empty}
\begin{table}
\begin{eqnarray*}
&\begin{tabular}{|c|c|ccc|cccccccc|cccccccc|}
\hline
 ~ & $\psi^\mu$ & $\chi^{12}$ & $\chi^{34}$ & $\chi^{56}$ &
        $\overline{\psi}^{1} $ &
        $\overline{\psi}^{2} $ &
        $\overline{\psi}^{3} $ &
        $\overline{\psi}^{4} $ &
        $\overline{\psi}^{5} $ &
        $\overline{\eta}^1 $&
        $\overline{\eta}^2 $&
        $\overline{\eta}^3 $&
        $\overline{\phi}^{1} $ &
        $\overline{\phi}^{2} $ &
        $\overline{\phi}^{3} $ &
        $\overline{\phi}^{4} $ &
        $\overline{\phi}^{5} $ &
        $\overline{\phi}^{6} $ &
        $\overline{\phi}^{7} $ &
        $\overline{\phi}^{8} $ \\
\hline
      $\alpha$ & 1 & 1 & 0 & 0 & 1 & 1 & 1 & 0 & 0 & 0 & 0 & 1 &
                 		 1 & 1 & 0 & 0 & 0 & 0 & 0 & 0 \\
       $\beta$ & 1 & 0 & 1 & 0 & 1 & 1 & 1 & 0 & 0 & 0 & 0 & 1 &
                                 1 & 1 & 0 & 0 & 0 & 0 & 0 & 0 \\
      $\gamma$ & 1 & 0 & 0 & 1 &
${1\over 2}$ & ${1\over 2}$ & ${1\over 2}$ & ${1\over 2}$ &
${1\over 2}$ & ${1\over 2}$ & ${1\over 2}$ & ${1\over 2}$ &
 0  & 0 & 0 & 0 & ${1\over 2}$ & ${1\over 2}$ & ${1\over 2}$ & ${1\over 2}$ \\
\hline
\end{tabular}
   \nonumber\\
&  ~ \nonumber\\
&  ~ \nonumber\\
&\begin{tabular}{|c|cccc|cccc|cccc|}
\hline
 ~&     $y^3y^6$ & $y^4\overline y^4$ & $y^5\overline y^5$ &
        $\overline y^3\overline y^6$ & $y^1\omega^6$ & $y^2\overline y^2$ &
        $\omega^5\overline \omega^5$ & $\overline y^1\overline\omega^6$ &
        $\omega^1\omega^3$ & $\omega^2\overline\omega^2$ &
        $\omega^4\overline\omega^4$ & $\overline\omega^1\overline\omega^3$ \\
\hline
       $\alpha$ & 1 & 0 & 0 & 0 & 0 & 0 & 1 & 0 & 0 & 0 & 1 & 1 \\
        $\beta$ & 0 & 0 & 1 & 0 & 1 & 0 & 0 & 0 & 0 & 1 & 0 & 0 \\
       $\gamma$ & 0 & 1 & 0 & 0 & 0 & 1 & 0 & 0 & 1 & 0 & 0 & 0 \\
\hline
\end{tabular}
\end{eqnarray*}
\caption{A three generation $SU(3)\times SU(2)\times U(1)^2$ model.
The choice of generalized GSO coefficients is:
$c{b_j\choose \alpha,\beta,\gamma}=
 c{\alpha\choose 1}=
 c{\alpha\choose \beta}=
 c{\beta\choose 1}=
 c{\gamma\choose 1}=
-c{\gamma\choose \alpha,\beta}=1~$
$(j=1,2,3)$, with the others
specified by modular invariance and space--time supersymmetry.
The sector ${\gamma}$ has the desired form $\gamma_L^2=\gamma_R^2=4$
and give rise to exotic diquarks.}
\label{modelwithexoticdiquarks}
\end{table}

\textwidth=7.5in
\oddsidemargin=-18mm
\topmargin=-5mm
\renewcommand{\baselinestretch}{1.3}
\begin{table}
\begin{eqnarray*}
&\begin{tabular}{|c|c|c|}
\hline
$ij$ & ${\lambda_{ij}^{'} \over x^n_a(i,j)} {100 GeV \over M_{LQ}} <$
      & Constraining Process \\
\hline
      33 & 0.03 & Charged Current Universality \\
     32 & 0.26 & Atomic Parity Violation and $eD$ Asymmetry \\
      31 & 0.26 & Atomic Parity Violation and $eD$ Asymmetry \\
      23 & 0.09 & $\Gamma(\pi \rightarrow e\nu)/\Gamma(\pi \rightarrow
      \mu\nu)$ \\
      22 & 0.22 & $\nu_{\mu}$ Deep Inelastic Scattering \\
      21 & 0.22 & $\nu_{\mu}$ Deep Inelastic Scattering \\
      13 & 0.84 & $\tau \rightarrow \pi \nu $ \\

\hline
\end{tabular}
\end{eqnarray*}
\caption{Experimental constraints on lepton number violating interactions
of exotic leptoquarks in models without baryon number violating operators
and without couplings of regular leptoquarks to standard model states.}
\label{lqtab}
\end{table}

\vfill
\eject

\begin{table}
\begin{eqnarray*}
\begin{tabular}{|c|c|c|rrrrrrrr|c|rr|}
\hline
  $F$ & SEC & $SU(3)_C\times SU(2)_L$&$Q_{C}$ & $Q_L$ & $Q_1$ &
   $Q_2$ & $Q_3$ & $Q_{4}$ & $Q_{5}$ & $Q_6$ &
   $SU(5)_H\times SU(3)_H$ & $Q_{7}$ & $Q_{8}$ \\
\hline
   $H_{13}$ & $b_1+b_3+$ & $(1,1)$ & $-{3\over4}$ & ${1\over2}$ &
   $-{1\over4}$ & ${1\over 4}$ & $-{1\over 4}$ & 0 & 0 & 0 &
   (1,3) & ${3\over4}$ & $5\over4$   \\
   $H_{14}$ &  $\alpha\pm\gamma+$  & (1,1) & ${3\over4}$  & $-{1\over2}$ &
   ${1\over4}$ & $-{1\over 4}$ & $1\over 4$ & 0 & 0 & 0 &
   (1,$\bar 3$) & $-{3\over 4}$ & $-{5\over 4}$  \\
   $H_{15}$ &     $(I)$       & (1,2) & $-{3\over4}$ & $-{1\over2}$ &
   $-{1\over4}$ & ${1\over 4}$ & $-{1\over 4}$ & 0 & 0 & 0 &
   (1,1) & $-{1\over 4}$ & $-{15\over 4}$  \\
   $H_{16}$ &               & (1,2) & ${3\over4}$ & $1\over2$ &
   ${1\over4}$ & $-{1\over 4}$ & $1\over 4$ & 0 & 0 & 0 &
   (1,1) & $1\over4$ & ${15\over 4}$  \\
   $H_{17}$ &               &  (1,1) & $-{3\over4}$ & $1\over2$ &
   $-{1\over 4}$ & $-{3\over4}$ & $-{1\over 4}$ & 0 & 0 & 0 &
   (1,1) & $-{1\over 4}$ & $-{15\over 4}$  \\
   $H_{18}$ &               & (1,1) & $3\over4$ & $-{1\over2}$ &
   ${1\over 4}$ & $3\over4$ & $1\over 4$ & 0 & 0 & 0 &
   (1,1) & ${1\over 4}$ & ${15\over 4}$  \\
\hline
   $H_{19}$ & $b_2+b_3+$ & (1,1) & $-{3\over4}$ & $1\over2$ &
   ${1\over 4}$ & $-{1\over4}$ & $-{1\over 4}$ & 0 & 0 & 0 &
   (5,1)&  $-{1\over 4}$ & ${9\over 4}$ \\
   $H_{20}$ & $\alpha\pm\gamma+$  & (1,1) & $3\over4$ & $-{1\over2}$ &
   $-{1\over 4}$ & $1\over4$ & $1\over 4$ & 0 & 0 & 0 &
   ($\bar 5$,1) & $1\over 4$ & $-{9\over 4}$  \\
   $H_{21}$ &  $(I)$         & (${\bar 3}$,1) & $1\over4$ & $1\over2$ &
   ${1\over 4}$ & $-{1\over 4}$ & $-{1\over4}$ & 0 & 0 & 0 &
   (1,1) & $-{1\over 4}$ & $-{15\over 4}$  \\
   $H_{22}$ &                & (3,1) & $-{1\over4}$ & $-{1\over2}$
   & $-{1\over 4}$ & ${1\over 4}$ & $1\over4$ & 0 & 0 & 0 &
   (1,1) & ${1\over 4}$ & ${15\over 4}$  \\
   $H_{23}$ &                & (1,1) & $-{3\over4}$ & $1\over2$ &
   ${1\over 4}$ & $-{1\over 4}$ & $3\over4$ & 0 & 0 & 0 &
   (1,1) & ${1\over 4}$ & ${15\over 4}$  \\
   $H_{24}$ &                & (1,1) & $3\over4$ & $-{1\over2}$ &
   $-{1\over4}$ & ${1\over 4}$ & $-{3\over4}$ & 0 & 0 & 0 &
   (1,1) & $-{1\over 4}$ & $-{15\over 4}$ \\
   $H_{25}$ &                & (1,1) & $-{3\over4}$ & $ 1\over2$ &
   $1\over4$ & $3\over4$ & $-{1\over4}$ & 0 & 0 & 0 &
   (1,1) & $-{1\over4}$ & $-{15\over4}$ \\
   $H_{26}$ &                & (1,1) & $3\over4$ &  $-{1\over2}$ &
   $-{1\over4}$ &  $-{3\over4}$ & $1\over4$ & 0 & 0 & 0  &
   (1,1) & $1\over4$ & $15\over4$  \\
\hline
\end{tabular}
\end{eqnarray*}
\caption{The exotic massless states from the sectors
$b_1+b_3+\alpha\pm\gamma+(I)$ and $b_2+b_3+\alpha\pm\gamma+(I)$,
in the model of Ref. \cite{eu}.}
\label{matter3}
\end{table}

\vfill
\eject

\begin{table}
\begin{eqnarray*}
\begin{tabular}{|c|c|c|rrrrrrrr|c|rr|}
\hline
  $F$ & SEC & $SU(3)_C\times SU(2)_L$&$Q_{C}$ & $Q_L$ & $Q_1$ &
   $Q_2$ & $Q_3$ & $Q_{4}$ & $Q_{5}$ & $Q_6$ &
   $SU(5)_H\times SU(3)_H$ & $Q_{7}$ & $Q_{8}$ \\
\hline
   $D_{1}$ & $b_2+b_3+$ & $({\bar 3},1)$ &
	${1\over4}$ & ${1\over2}$ &
   ${1\over4}$ & $-{1\over 4}$ & $-{1\over 4}$ & 0 & 0 & 0 &
   (1,1) & $-{1\over4}$ & $-{{15}\over4}$   \\
   ${\bar D}_{1}$ & $\beta\pm\gamma+$  & $(3,1)$ &
	$-{1\over4}$  & $-{1\over2}$ &
   $-{1\over4}$ & ${1\over 4}$ & ${1\over 4}$ & 0 & 0 & 0 &
   (1,$1$) & ${1\over 4}$ & ${{15}\over 4}$  \\
\hline
   $D_{2}$ & $b_1+b_3+$ & $({\bar 3},1)$ &
	${1\over4}$ & ${1\over2}$ &
   $-{1\over4}$ & ${1\over 4}$ & $-{1\over 4}$ & 0 & 0 & 0 &
   (1,1) & $-{1\over4}$ & $-{{15}\over4}$   \\
   ${\bar D}_{2}$ & $\alpha\pm\gamma+$  & $(3,1)$ &
	$-{1\over4}$  & $-{1\over2}$ &
   ${1\over4}$ & $-{1\over 4}$ & $1\over 4$ & 0 & 0 & 0 &
   (1,$1$) & ${1\over 4}$ & ${{15}\over 4}$  \\
\hline
   $H_{1}$ & $b_2+b_3+$ & $(1,1)$ &
	$-{3\over4}$ & ${1\over2}$ &
   $-{1\over4}$ & ${1\over 4}$ & $-{1\over 4}$ & 0 & 0 & 0 &
   (1,3) & ${3\over4}$ & ${{5}\over4}$   \\
   ${\bar H}_{1}$ & $\beta\pm\gamma+$  & $({1},1)$ &
	${3\over4}$  & $-{1\over2}$ &
   ${1\over4}$ & $-{1\over 4}$ & $1\over 4$ & 0 & 0 & 0 &
   (1,${\bar 3}$) & $-{3\over 4}$ & $-{{5}\over 4}$  \\
\hline
   $H_{2}$ & $b_2+b_3+$ & $(1,1)$ &
	$-{3\over4}$ & ${1\over2}$ &
   ${1\over4}$ & $-{1\over 4}$ & $-{1\over 4}$ & 0 & 0 & 0 &
   (1,3) & ${3\over4}$ & ${{5}\over4}$   \\
   ${\bar H}_{2}$ & $\alpha\pm\gamma+$  & $({1},1)$ &
	${3\over4}$  & $-{1\over2}$ &
   $-{1\over4}$ & ${1\over 4}$ & $1\over 4$ & 0 & 0 & 0 &
   (1,${\bar 3}$) & $-{3\over 4}$ & $-{{5}\over 4}$  \\
\hline
\end{tabular}
\end{eqnarray*}
\caption{The exotic massless states from the sectors
$b_1+b_3+\alpha\pm\gamma+(I)$ and $b_2+b_3+\alpha\pm\gamma+(I)$,
in the model of Ref. \cite{gcu}.}
\label{matter4}
\end{table}

\end{document}